# Seasonal Changes – Time for Paradigm Shift

Branislava Lalic and Ana Firanj Sremac

Group for Meteorology and Biophysics, Faculty of Agriculture
University of Novi Sad, Novi Sad, Serbia

**Abstract**

Season and their transition play a critical role in sharpening ecosystems and human activities, yet their traditional classifications – meteorological and astronomical – fail to capture the complexities of biosphere-atmosphere interactions. The conventional definitions often overlook the interplay between climate variables, biosphere processes (including human activities), and the actual anticipation of seasons, particularly in the context of global climate change, which has disrupted the traditional seasonal patterns.

This study addresses the limitations of current seasonal classification by proposing a framework based on phenological markers such as Normalized Difference Vegetation Index (NDVI), Enhanced Vegetation Index (EVI), Leaf Area Index (LAI), Fraction of Photosynthetically Active Radiation (fPAR), and the Bowen ratio, using plants as a nature-based sensor of seasonal transitions. These indicators, derived from satellite data and ground observations, provide robust foundations for defining seasonal boundaries. We hypothesize that the normalized daily temperature range (DTRT), already validated in crop and orchard regions, serves as a reliable, biologically relevant **seasonality index** capable of capturing seasonal transitions.

We demonstrated that the seasonality index aligns closely with phenological markers and ground observations across diverse climatic regions, including boreal, temperate, and deciduous forests. By analyzing trends, extreme values and inflection points in the seasonality index time series, the study establishes a methodology for identifying the onset, duration, and transition of seasons. The new classification offers universal and scalable metrics that align with current knowledge and perception of seasonal shifts while effectively capturing site-specific timing and duration of seasons. Findings based on this new classification reveal significant shifts in seasonal patterns across the Euro-Mediterranean region, with winters shortening, summers extending, and transitions becoming more pronounced. Notable specific effects include the Gulf Stream's influence on milder seasonal transitions, the urban heat island effects of large cities accelerating spring and summer transitions, and the impact of large inland lakes moderating seasonal durations.

This underscores the critical importance of deeper knowledge about seasonal transitions, which enables informed decision-making in everyday life and professional sectors such as agriculture, forestry, urban planning, medicine, trade, marketing, tourism, and even the stock market. These changes reflect broader climate trends, emphasizing the urgency for adaptive strategies in resource management and policy-making.

**Keywords:** definition of seasons, seasonal classification, phenological markers, seasonal shifts, normalized daily temperature range

## 1 Introduction

Seasons and seasonal changes are complex phenomena. Their quantitative description is typically related to the magnitude, frequency, and phase of climate variables that describe the state of the atmosphere (Stine, et al., 2009 Metzger, et al., 2005). Qualitatively, seasons are often associated with biosphere processes such as a plant (Kwiecien et al., 2022) and insect phenology (Danks 2007, Dümpelmann 2025), bird migration (Jenni and Kéry, 2003, Van Buskirk et al., 2009), and human comfort (Graff Zivin and Neidell 2014, Cohen et al., 2012, Hedquist and Brazel, 2014). However, seasonal transitions and their duration also impact human health, emotional states, and social interactions (Rintamäki et al., 2008, Wehr et al., 2001, Coletti et al., 2018, Obradovich et al., 2018, Harp and Karnauskas, 2018). From an economic and societal perspective, shifts in season duration and characteristics influence agriculture, tourism, and other businesses (Lobell et al., 2011, Scott et al., 2012).

The paradigm science follows regarding seasons offers two, broadly accepted, definitions of seasons: meteorological and astronomical. Whereas the meteorological seasons are based on the annual temperature cycle, lasting exactly three months, astronomical seasons are based on the Earth's position as it rotates around the sun. The existing paradigm was criticized by scientists from the early 1980ies (Trenberth, 1983), particularly due to its inability to account for the complex interactions between climate variables and biological processes, especially in light of global climate change (CC) (Peñuelas and Filella, 2001, Cassou and Cattiaux, 2016, Kirbyshire and Bigg, 2010). Current CC completely disturbed seasonal clock (Schwartz et al., 2006, Crimmins and Crimmins, 2019, Thomson 2009, Sparks and Menzel, 2002), raising awareness that the concept of the four seasons of equal length in mid latitudes and sub-polar regions should be dismissed (Wang et al., 2021).

Recent studies (Wang et al., 2021; Littleboy et al., 2024) have attempted alternative approaches, but many fall short in providing scientifically sound solutions that are applicable in daily life. Therefore, developing a seasonal classification that accurately monitors and forecasts seasonal transitions is of utmost importance. To address the limitations of existing definitions of seasons, we propose a new classification based on natural phenomena, particularly focusing on plant dynamics. This allows for accurate monitoring and forecasting of seasonal transition, onset, and duration. We will suppose that seasons can be defined based on natural phenomena, primarily plant dynamics. We will call them 'seasonal phenomena'. Timing and characteristic values of the extremes and inflection points in climate data and derived indices time series, which can be associated with the transition of seasons, shall be referred to as seasonality indices. We hypothesize that, at a given location, seasonality indices correspond to the seasonal phenomena. Defining a seasonality index that encapsulates seasonal transitions using only meteorological variables offers a way to predict and track these changes regardless of plant presence.

In our study, we think of plants as biological sensors, reacting to changes in their environment much like how litmus paper reacts to pH. If plants "turn green" (beginning of growing season) or "turn brown" (start of dormancy) based on atmospheric conditions, then phenomena that best depict these transitions—such as leaf area development, changes in leaf coloring, and fluctuations in physiological processes like evapotranspiration—are likely to be key indicators of seasonal shifts. Plant changes in size, expressed through Leaf Area Index (LAI) (Parker 2020, Barr et al., 2004), and color, commonly measured using Normalized Difference Vegetation Index (NDVI) and Landsat Enhanced Vegetation Index (EVI) indices (Huete et al., 2002, Soudani et al., 2012) are some of the best documented remotely sensed markers of growing seasons (Reed et al., 2003, Wu et al., 2017, Steltzer and Post, 2009). On the other hand, changes in the intensity of physiological processes are clearly depicted by seasonal variation of atmospheric indices such as a fraction of incoming PAR (fPAR) (Cheng et al., 2014) and Bowen ratio (Fitzjarrald, 2001; Figs. 1 and 2).

We hypothesize that a more accurate and flexible classification of seasons can be achieved by using listed bio-based markers and atmospheric indices. Based on this reasoning we adopted the following description of seasons.

- Spring begins with the onset of the growing season, characterized by increasing photosynthesis, evapotranspiration, and leaf area development.

- Summer represents a period of biological maturity and stability in ecosystems, where energy is focused on maintaining growth and productivity. The constant green area in forests and active yield formation in crop production are valid proxies for summertime description.

- Autumn starts as plants enter dormancy, marked by reduced evapotranspiration and physiological activity and takes as long as perennial plants hit full dormancy.

- Winter is the period of minimal activity and metabolic slowdown, as ecosystems reach full dormancy.

To capture these transitions quantitatively, we use indices closely related to plant phenology (NDVI, EVI, LAI, fPAR, Bowen ratio - in further text referred to as 'phenology markers') and compared their annual pattern with the normalized Daily Temperature Range (DTRT = ($T_{max}$-$T_{min}$)/$T_{avg}$) pattern to test its capacity to act as a seasonality index. Based on its demonstrated ability to accurately depict seasonal transitions in the crop-growing Pannonian region (Lalic et al., 2022), we hypothesize that DTRT can serve as a reliable, biologically relevant metric for defining seasons, independent of pre-defined temperature thresholds. To test this hypothesis, we analyze extreme values and trends in phenology markers alongside DTRT patterns. This approach establishes a robust methodology for defining seasonal boundaries (Sec. 2.2) while also validating DTRT as a reliable seasonality index (Sec. 3.1). To put our new season classification into force, we calculated the duration of seasons over Euro-Mediterranean region (Sec. 3.2). Instead of comparing the onset of seasons between two climatological periods, we made a comparison among decades during the last climatological period. The reasoning for this is twofold: i) numerous studies have already established that there is a significant shift in seasons between current and past climatology (see for example, Twardosz et al., 2021; Jylha et al., 2010; Ruosteenoja and Raisanen, 2013; Kuglitsch et al., 2010; Zorita et al., 2010) and ii) all millennial records in temperature happened during the last climatological period. This period we considered particularly important to address the high rate of warming in and its effect on the onset and duration of seasons.

*Rationale behind use of DTRT*. First we tested daily temperature range as an index often used in plant phenological studies (Huang et al., 2020, Wang and Liu, 2023). However, its sensitivity to several factors unrelated to plant phenology or seasonal transition, such as cloud cover, humidity, and extreme weather events such as heat waves and cold spells was a major limitation (Hanes, 2014, Cox et al., 2020). Comparison to listed indices, DTRT is a simpler, more universally applicable metric that captures the relative variability of daily temperature. By normalizing the temperature range by the average temperature, the index provides a dimensionless value that is more applicable across different climate zones. Its key advantage is that it describes intrinsic seasonal dynamics of the atmosphere with or without vegetation. This makes it particularly valuable in areas where plant life is sparse or absent, while still effectively reflecting seasonal changes. The DTRT index cannot be calculated when the daily temperature is 0 °C. However, this limitation is negligible for our purposes, as our focus lies primarily on extreme values and inflection points, which occur outside of winter conditions and remain unaffected by this constraint.

**2 Methodology**

2.1 Data

This study utilizes in-situ micrometeorological and phenological data as well as satellite-derived data collected from three forest sites, each representing a specific climatic region:

- **Canada (CA)**: CA-Oas, Saskatchewan - Western Boreal, Mature Aspen forest, located at 53.62889° N, 106.19779° W, with an elevation of 600.63 m (boreal forest); Köppen-Geiger (KG) classification: Dfc (Subarctic: severe winter, no dry season, cool summer) (Black, 2016).
- **Germany (DE)**: DE-Hai, Hainich National Park, at 51.0792° N, 10.4522° E, with an elevation of 430 m, characterized by a mixed deciduous beech forest (temperate forest); KG classification: Cfb (Temperate oceanic: warm summer, no dry season) (Knohl et al., 2016).
- **United States (US)**: US-PFa, Park Falls/WLEF, located at 42.537755° N, 72.171478° W, with an elevation of 470 m, representing a mixed deciduous forest (deciduous forest); KG classification: Dfb (Warm Summer Continental: significant precipitation in all seasons) (Desai, 2016).

*2.1.1 Ground-based micrometeorological data and calculated indices*

Micrometeorological data for the calculation of the **Bowen Ratio** and **DTRT** were sourced from the FLUXNET database (http://fluxnet.fluxdata.org/). FLUXNET provides continuous, high-frequency measurements of energy, water, and carbon fluxes, which support the calculation of energy partitioning and vegetation seasonality (Pastorello et al., 2020). The **Bowen Ratio** was calculated using sensible and latent heat flux while for DTRT index we used tower measurements of air temperature above forest vegetation. Both flux variables were gap-filled using the Marginal Distribution Sampling (MDS) method while the air temperature is gap-filled by MDS and ERA-Interim reanalysis by FLUXNET as described in Pastorello et al., 2020.

To provide a comprehensive analysis of season duration based on DTRT annual cycle over Europe we used ERA5-Land reanalysis data (Copernicus Climate Change Service, 2019) for 1991-2020.

*2.1.2. Satellite-derived data*

Satellite-derived data were obtained from the Terra Moderate Resolution Imaging Spectroradiometer (MODIS).

The **MOD13A1** product provides 16-day composites at 500 m resolution, including **NDVI** and **EVI**, which were extracted based on pixel quality criteria, such as low cloud cover, optimal view angles, and the highest index values covering the period from 18 February 2000 to the present (Didan, 2021). The **MOD15A2H** Version 6.1 product, covering the same period, provides 8-day composites of **LAI** and **fPAR** at a 500 m resolution, with the best available pixel selected for each period. **LAI** represents the green leaf area per unit ground area, while **fPAR** measures the fraction of incoming PAR (400-700 nm) absorbed by the green vegetation canopy. This dataset also includes additional quality layers and standard deviation metrics to support detailed analysis of canopy structure and light absorption (Myneni et al., 2021).

*2.1.3 Ground-based phenological data*

For phenological data in this study we used narrative description and BBCH ("Biologische Bundesanstalt, Bundessortenamt und CHemischeIndustrie") scale (Meier, 2018) for phenological phases denotation, while associated date is expressed as a day of the year (DOY)

**DE**. For the DE location, phenological data were sourced from the PEP725 database (Templ et al., 2018, Data set accessed 2024-10-11 at http://www.pep725.eu), focusing on all beech phenological observations conducted within the last 30 years, in a 30 km radius of the tower and for 5 consecutive years. After filtering, only observations of three BBCH stages remained: BBCH 11 (leaf unfolding, first visible leaf stalk) with 368 station-years, BBCH 94 (50% leaves discolored) with 360 station-years, and BBCH 95 (50% of leaves fallen) with 325 station-years. The average occurrence dates for these stages were 115 DOY, 282 DOY, and 299 DOY, respectively.

**US**. Since 1990, ground-based phenological observations of woody vegetation at Harvard Forest have documented critical stages of development. Initially covering 33 species, the study focused on bud break and leaf development, later narrowing to nine representative species for spring and 14 for fall phenology in 2002 (O'Keefe and VanScoy, 2024, Data set accessed 2024-10-15 at https://harvardforest.fas.harvard.edu/). Four key phenological events were recorded per species: 50% buds open with visible leaves (BBCH 11), 50% of leaves reaching 75% of final size (BBCH 19), 50% of leaves discolored (BBCH 94), and 50% of leaves fallen (BBCH 95). On average, bud break occurs on day 125 DOY, 75% leaf development on day 148 DOY, leaf discoloration on day 281 DOY, and leaf fall on day 291 DOY.

**CA**. The phenological observations for CA were from the PhenoCam , of green chromatic coordinate (Gcc) (Hufkens et al., 2018, Seyednasrollah et al., 2019, Data set accessed 2024-10-24 at https://daac.ornl.gov/) and literature: start of spring green-up – 129 +/- 12 DOY, 90% aspen leaf emergence – 153 +/-7 DOY, end of green-up: 162 +/- 8 DOY, end of senescence – 278 +/-7 DOY in average for 1994-2003 (Barr et al., 2004). Additionally, from Blanken et al. (1997) we find out that in this forest leaf growth began during the third week of May (135 DOY) with a maximum forest leaf area index of 5.6 $m^2$ $m^{-2}$ reached by mid-July (200 DOY).

2.2 Data Processing and Analysis

For this study, several key variables were extracted and processed.

The Bowen Ratio (B) was derived using the ratio of hourly sensible (H_CORR) and latent heat flux (LE_CORR) from the FLUXNET database (Pastorello et al., 2020). Quality checks were conducted on these variables using the quality flags included in the FLUXNET dataset to ensure an accurate representation of energy partitioning at the selected sites. Only mean values for the midday period (B_mmd), calculated as the average between 10:00 and 14:00, are used.

Satellite data were processed using Google Earth Engine for consistency across all sites. Data extraction was conducted at specific site coordinates, aligning with the 500 m resolution of MODIS pixels. High-quality data were retained for LAI, fPAR, NDVI, and EVI using quality control bands, and values were scaled according to product specifications. Temporal alignment was achieved by averaging data over intervals matching the in-situ measurements, ensuring comparability.

A smoothing methodology is applied to interpolate and handle sparse data, such as satellite observations. For this purpose, we implemented a Generalized Additive Model (GAM) using the R package "mgcv" (Wood, 2011, 2017). The procedure follows the approach described in Lalic et al. (2022).

To harmonize ground phenological data, we utilized the BBCH scale as provided in the PEP725 dataset. Data from two North American sites were converted to the BBCH format based on the methodology outlined by Finn et al. (2007) and given in PEP725 database.

We calculated seasonality index DTRT using data from two sources. The first source was air temperature measurements from flux towers provided by FLUXNET, where we utilized hourly data to calculate the maximum, minimum, and average values (TA_F). Not all sites provided the same time series period; for example, CA covered 1996–2010, DE 2000–2012, and US 1994–2014. The second source was the ERA5-Land reanalysis data. From this dataset, we extracted hourly temperatures for the period covered by flux measurements at each specific location and for the nearest grid point. From hourly 2m air temperatures we calculated the maximum, minimum, and average values required for DTRT index.

2.3 Definition of seasons

Annual variations in phenology markers and DTRT index (Fig. 1), exhibit distinct seasonal patterns that enable the identification of season boundaries. These patterns are consistent across diverse geographical locations (Canada, Germany, and the United States) and emphasize periods of near-constant values during winter and summer, coupled with linear transitions during spring and autumn. By integrating these trends, we developed a methodology for determining onset and duration of seasons, as well as the start (SOS) and end (EOS) of growing season, based on timing of **extreme values and inflection points** of DTRT index. The methodology is illustrated on Figure 2 using satellite LAI annual variations on CA location.

**Winter** is characterized by constant values of phenological markers, except for high variability of Bowen ratio. The winter end (WE) corresponds to the first minimum in satellite indices time series and the first maximum in DTRT time series. It signals the beginning of spring.

**Spring** is defined by a steady rise in satellite indices and corresponding linear decrease in Bowen ratio and DTRT index. SOS occurring in this period is typically set on 30% of LAI annual maximum (Bórnez et al., 2020, Verger et al., 2016) or 50% of NDVI or EVI based on dynamic threshold method (Reed et al., 2003). According to ground observations and previous experience with crops and orchards (Lalic et al., 2022), SOS in DTRT time series is set at the first inflection point at approximately 50% of its annual amplitude. The duration of spring is bounded by the regression periods of both phenology markers and DTRT index, highlighting their complementary trends in marking the transition to summer.

**Summer** period aligns with peak of vegetation activity, as indicated by maximum values of LAI, NDVI and fPAR and minimum values of Bowen ratio (when latent heat flux overcomes sensible) and DTRT index. According to (Wang et al., 2017) summer corresponds with period when LAI values exceed 80% of the annual maximum, and it perfectly fits period of constant DTRT values on all locations (summer start (SS)-end (SE)). The duration of summer is defined as period of constant values of phenology markers and DTRT index.

**Autumn** is characterized by a linear decrease in satellite indices, accompanied by a simultaneous linear increase in Bowen ratio and DTRT index, representing the gradual transition from summer to winter. The end of autumn, i.e. autumn-winter transition is marked by increased variability of Bowen ratio (heading towards winter trend) and sudden drop of DTRT index signaling the onset of winter (WS). EOS occurring in this period is typically set on 40% of LAI annual maximum (Bórnez et al., 2020, Verger et al., 2016) or 50% of NDVI or EVI based on dynamic threshold method (Reed et al., 2003), and it corresponds with DTRT second inflection point at, again, 50% of its annual amplitude.

**Transition period (TR)** between the seasons depicts a period between two distinct trends; highlights the gradual nature of seasonality and offers possible insight into the dynamics of phenological changes.

**Classification summary.** According to the obtained results, in further analysis, DTRT-based seasons will be defined as follows: WE – first DTRT maximum; spring – after WE, period of DTRT linear decrease; summer–-period of DTRT constant values including transition periods

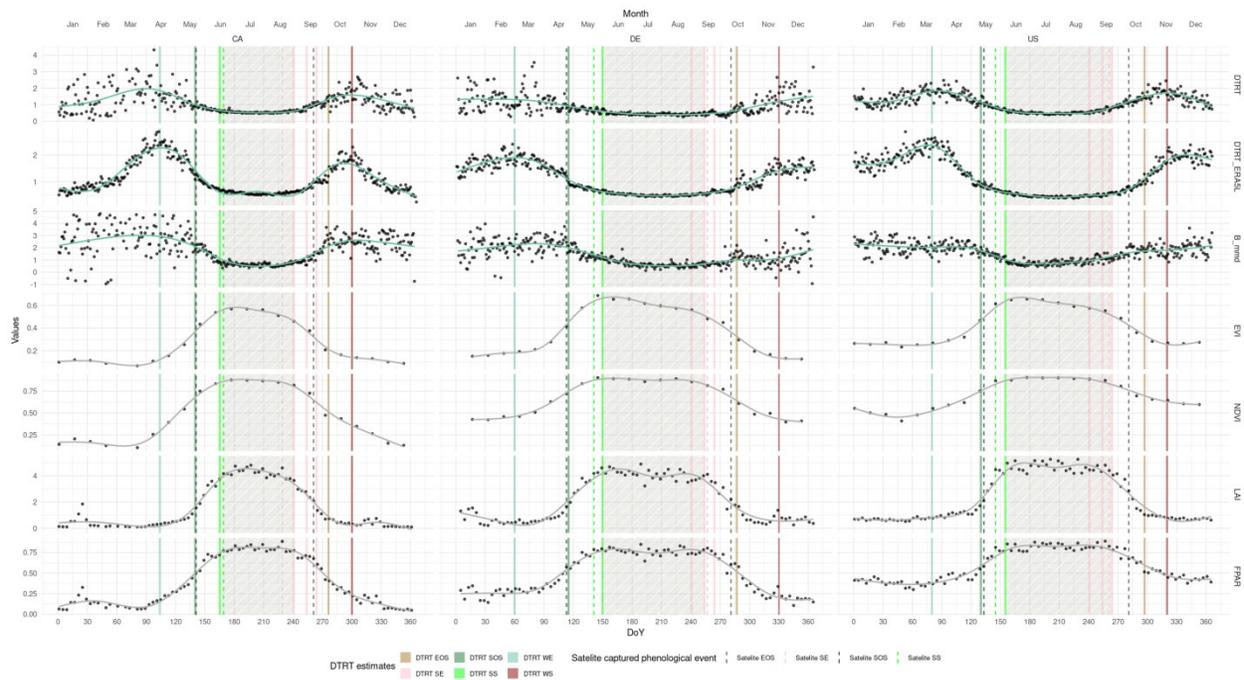

**Figure 1** Long term averages of NDVI, EVI, LAI, fPAR (maximum), mean midday Bowen ratio (B_mmd) and DTRT in CA, DE, US averaged for period covering data availability.

(SS-SE); autumn – period of DTRT linear increase; WS – period of DTRT constant values at the end of the year; SOS and EOS – first and second inflection points of DTRT time series, corresponding to approximately 50% of annual amplitude.

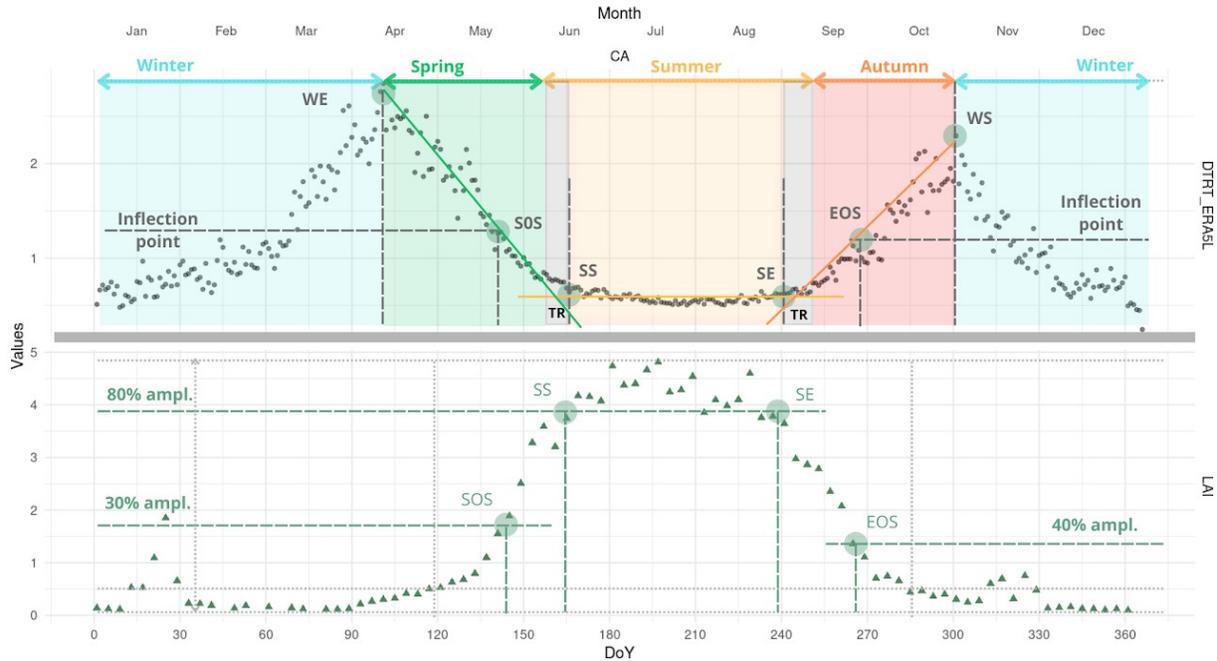

**Figure 2** Graphical presentation of methodology used to identify duration of seasons on CA location

## 3 Results

3.1 Seasonality index

To validate DTRT as a seasonality index we compared satellite-derived timing of phenological events and ground observations associated with season transitions with DTRT-based results (Fig. 1 and Tab. 1).

**Winter.** The first annual maximum of the DTRT index determines the winter end (WE). It corresponds with the first annual minimum and slight start of satellite indices increase (DE and USA), particularly NDVI and EVI primarily driven by grass greening and spring ephemerals. This transition signals the onset of spring. At US location WE typically appear at **80** DOY. It corresponds with results obtained by Fitzjarrald et al. (2001) for early spring at **90** DOY.

**Spring.** DTRT-based spring is the regression period of both phenology markers and the DTRT index highlighting their complementary trends in marking the transition to summer. Ground observations capture vegetative development, reflected in BBCH stages 11-19. On all locations, DTRT-based SOS deviates 1-3 days from satellite-based SOS, and 5-11 days from ground truth, i.e. BBCH stage adopted as SOS. In US forest, SOS determined as DTRT inflection point at **130** DOY corresponds with Fitzjarrald's spring date, determined at DOY **125** as time of bud break.

**Summer.** The duration of summer derived from the DTRT- and satellite-based indices, is completely aligned on all locations, with the exception of summer end based on LAI in CA, where outliers at the end of August affect the result. However, summer end based on NDVI and EVI trends correspond with DTRT-based results. Ground observations of BBCH stages 19-92, reflecting full vegetation growth and ripening, refer to period that starts earlier and takes longer than it is expected from DTRT- and satellite-based assessment of summer duration.

**Autumn.** Early autumn, just after the summer end, is characterized by an intensive decrease of all satellite-based indices and an increase of the DTRT index reaching its second annual maximum, approximately at the end of October. After that time, all indices decrease marking the end of autumn and the onset of winter (WS), defined by return of LAI and DTRT to their respective annual extremes. Particularly interesting is a transition period between autumn and winter depicted by DTRT trends sliding between autumn linear regression and winter constant values, highlighting the utility of DTRT in marking seasonal transitions. Ground observations capture vegetative senescence and leaf fall (BBCH: 92-94). DTRT- and satellite-based timing of EOS are aligned on DE location while sudden drop in LAI on CA and USA locations brings two weeks difference.

| Location | Winter end (WE) | Veg. start (SOS) | Summer start (SS) | Summer end (SE) | Veg. end (EOS) | Winter start (WS) | Index |
|---|---|---|---|---|---|---|---|
| Ground truth | | BBCH 11 | BBCH 19 | | BBCH 95 | | |
| CA | - | 129 | 153 | - | 278 | - | Ground truth |
| | 109/60 | 140 | 165 | 241 | 276 | 328 | DTRT |
| | - | 141 | 169 | 229 | 261 | - | LAI |
| DE | - | 115 | - | 282 | 299 | - | Ground truth |
| | 78/40 | 115 | 150 | 254 | 287 | 330 | DTRT |
| | - | 113 | 141 | 257 | 281 | - | LAI |
| US | | 125 | 148 | 281 | 291 | - | Ground truth |
| | 90/72 | 130 | 155 | 264 | 297 | 345 | DTRT |
| | - | 133 | 145 | 261 | 281 | - | LAI |

**Table 1** Timing of the onset of seasons, start (SOS) and end (EOS) of growing season obtained using DTRT extreme values and inflection points, threshold values of LAI and ground observations of associated phenological stages

3.2 Onset and duration of seasons

The new classification of seasons and derived metrics for DTRT as a seasonality index we applied on ERA5-Land reanalysis data to obtain new insights into the Euro-Mediterranean region's seasonal dynamics shaped by regional and local climate and geography (Figs. 3-8).

The longer durations of **winter** in northern and northeastern Europe (Fig. 3), with durations exceeding 200 days in regions like Scandinavia and parts of Eastern Europe are consistent with extended cold periods and minimum vegetation activity. Moving southward, winters grow shorter reflecting milder climates of southern (S) Europe and

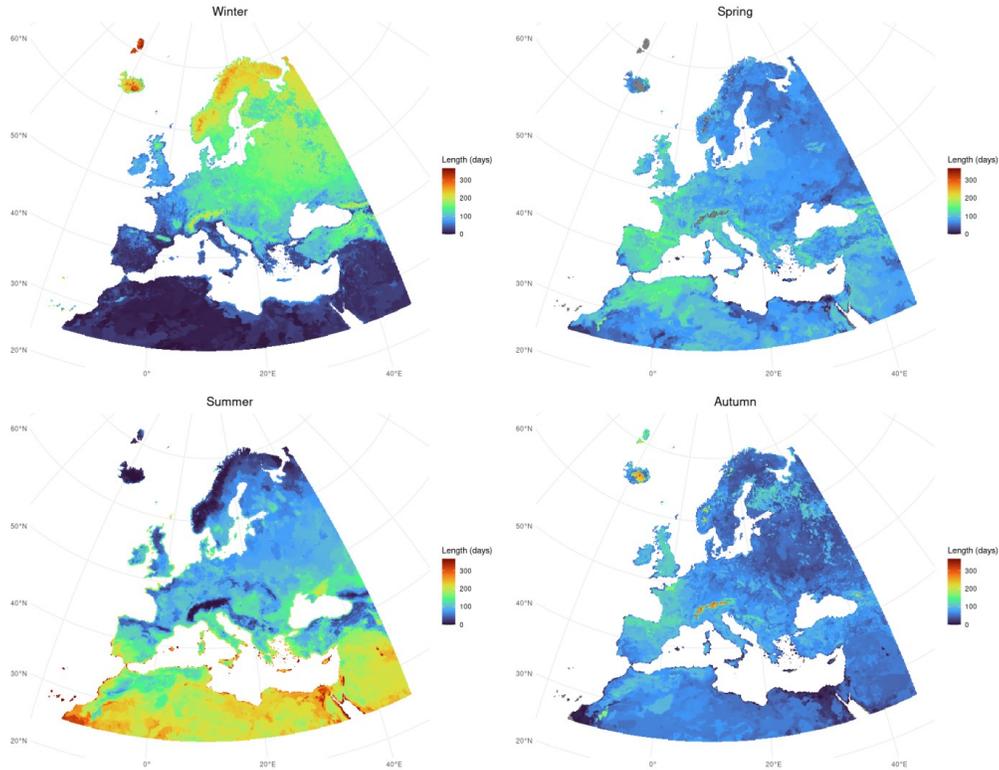

**Figure 3** Duration of seasons over the Euro-Mediterranean region for 1991-2020 ERA5 Land climatology

Mediterranean region experiences much shorter winters, often lasting less than 100 days. Exceptions are high mountain regions Pyrenees (Spain), the Atlas Mountains (northwestern Africa) and the Pindus Mountains (Greece) whose longer winter period is depicted on the map. **Spring** is most pronounced in mid-latitudes, where central Europe experiences spring duration ranging between 50 and 100 days. However, moving northward, spring becomes compressed due to prolonged winters, while in the south, the transition to summer occurs more abruptly. **Summer**, in turn, dominates the seasonal calendar in the Mediterranean region, where it lasts more than 200 days, far surpassing its duration in northern Europe, where it rarely exceeds 100 days. **Autumn**, much like spring, exhibits significant variability, with the longest duration in mid-latitudes. In northern Europe, autumn transitions are sharp, giving way to early winters, while in southern Europe autumn remains constrained by the extended presence of summer. Well depicted are specifics of **weather in Ireland and Great Britain** islands where shorter winter and summer times are replaced with extended spring and autumn periods. An exception is Scotland with it's highlands exerting longer winter and shorter summer times. **Gulf Stream** effects can be seen in the form of shorter winter and longer spring times on the west coast of Europe up to Scandinavia, particularly on the southern and southwestern coast of Norway and the west coast of Sweden (shorter winter) and overall mild maritime climate (shorter winter + summer; longer spring + autumn) typical for south-southwestern part of Sweden and most of Denmark. Landlocked water bodies such as Russia's Ladoga and Onega lakes, Sweden's Vänern Lake, and the Sea of Azov are distinctly visible on the season duration maps. These water bodies moderate local climates, resulting in longer summers and shorter winters compared to the surrounding regions. The extended duration of autumn in the mountain area of central and southern Europe is on account of the shorter summertime. The places of constant snow and ice, such as the peaks of the Alps, Scandinavia, and Iceland, demonstrate a distinct seasonal pattern characterized by only two predominant seasons: a warm season (autumn) and a cold season (winter). In these high-altitude and high-latitude regions, the warm season aligns with the period of minimal snow cover and slightly elevated temperatures, while the cold season dominates due to persistent snow and ice cover and extreme cold conditions. This binary seasonal classification reflects the limited vegetation activity and harsh climatic constraints in these areas. When comparing seasonal durations across three decades from 1991-2000 till 2011-2020, clear shifts emerge (Fig. 4). Winter duration has shortened across most of the Euro-Mediterranean region, with some regions seeing reductions of more than 30 days per decade. Spring and autumn have extended in many areas on winter account. Shortening of spring is associated with regions with earlier summer start and typically longer summers. Summer (with included transition periods) shows significant expansions, with durations increasing by more than 30 days per decade in southern and central Europe and the Mediterranean region between the first and last decade. These results align with warming trends and a delayed transition to cooler seasons. The autumn trend varies. The shortening of autumn in the Mediterranean region and eastern Europe is a result of prolonged summer, while in some parts of northern, western, and eastern Europe autumn is prolonged due to shorter summer and later start of winter.

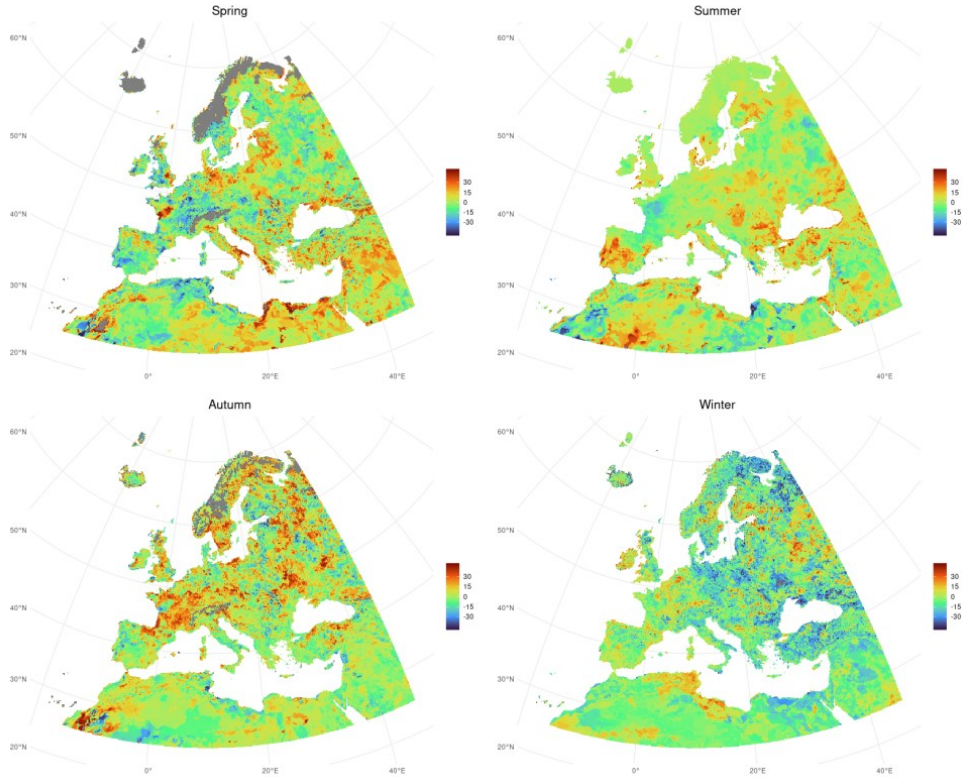

**Figure 4** Shifts in seasonal duration over the Euro-Mediterranean region for 1991-2020 ERA5-Land climatology between first (1991-2000) and last (2010-2020) decades.

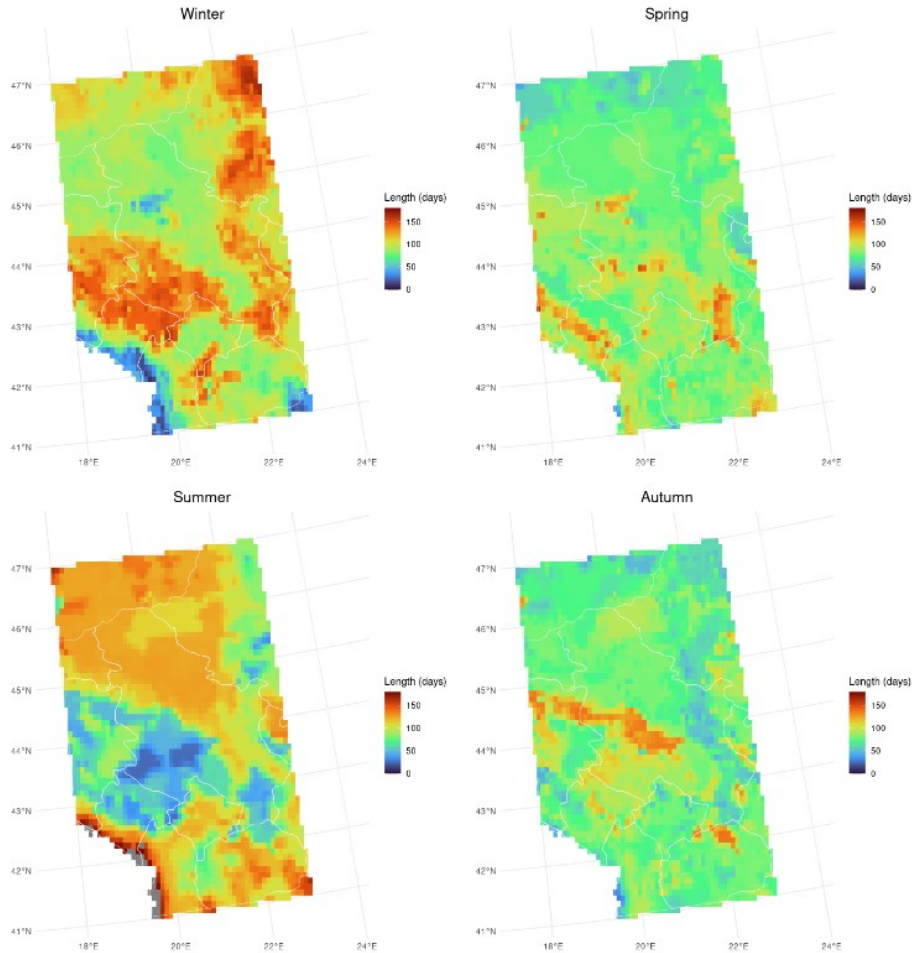

**Figure 5** Duration of seasons over the SE Europe for 1991-2020 ERA5-Land climatology

A closer look on a regional scale (Fig. 5) brings fine distinction in winter duration on mountains of different heights and deeply continental northern part of the region, in comparison to southwestern and southern areas which are under a strong influence of warmer air coming, in this period, from Adriatic Sea (southwest) and Aegean Sea through Vardar river valley (south). Urban heat islands in Novi Sad and Belgrade cities (approximately at 45 °N) are depicted with shorter winters and longer springs. Extensive summer periods in the Pannonian lowlands are seen in the northern part of the region and narrow coastal areas of Bosnia and Herzegovina, Montenegro and Albania.

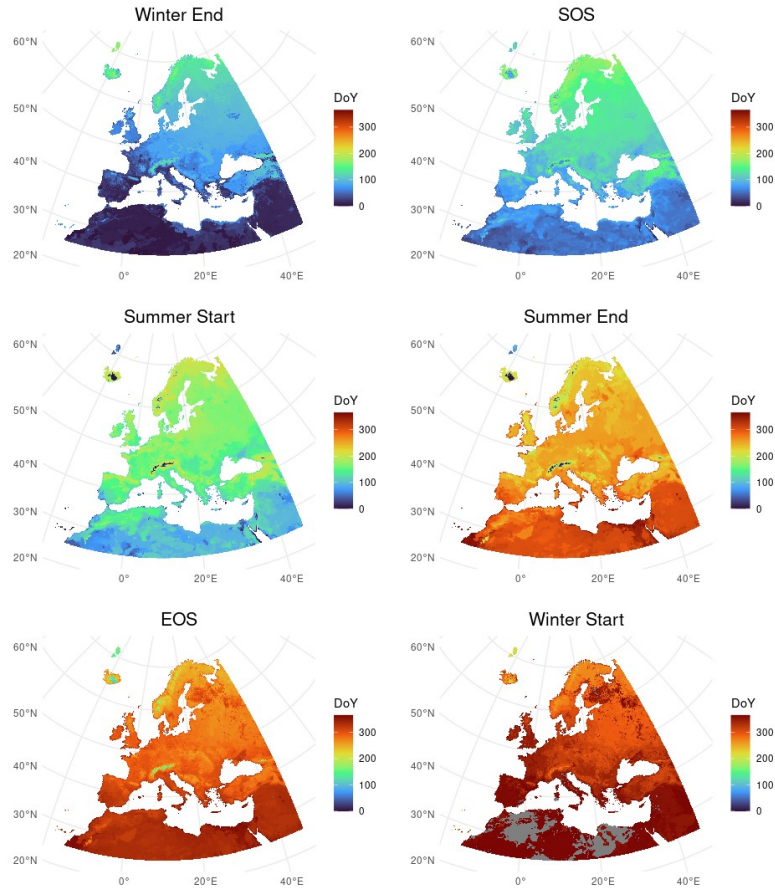

**Figure 6** Onset of seasons over the Euro-Mediterranean region for 1991-2020 ERA5-Land climatology

The timing of seasonal transitions across the Euro-Mediterranean region (Fig. 6) shows distinct spatial and temporal patterns influenced by latitude, altitude, the proximity of water bodies, and continentality. Winter ends earlier in coastal and southern regions, particularly along Atlantic and Mediterranean coasts, where warmer air in the winter accelerates the seasonal transition. Inland and northern regions, such as Scandinavia and Russia, as well as high mountain regions like Alps, Dinarides, Carpathian, and, even Atlas Mountains experience a delayed end of winter. Landlocked water bodies contribute to earlier transitions by moderating local climates. The superiority of DTRT as a seasonality index able to depict small-scale differences in season transitions can be seen from differences in winter duration and onset obtained for Canary Islands and nearby coastal areas of Western Africa. These differences arise due to the Islands unique geographical location, diverse landscape and elevations and climate strongly influenced by Atlantic Ocean and trade winds, while climate of Western Africa even on the coast is mainly affected by Sahara Desert. The start of the growing season (SOS) closely follows the end of winter. Spring begins earliest in southern Europe and Mediterranean regions where spring transitions start at approximately **60–80** DOY. Northern and eastern Europe show later spring onsets, often exceeding **120** DOY, reflecting colder and more prolonged winters. The urban heat island effect associated with Moscow, Russia (55.75°N, 37.62°E), is visible on the winter end and SOS maps. It indicates an earlier end of winter and earlier SOS in the area east of the city. This pattern is likely influenced by the prevailing western and southwestern winds during winter in this region.

The differences in seasonal transitions during three decades of 1991-2020 climatology reveal substantial changes in the timing of seasons that reflect broader regional climate trends (Fig. 7). These shifts are characterized by pronounced spatial variability, highlighting the heterogeneous impact of climate change.

Over the last three decades, the onset of the SOS has advanced significantly in large parts of central and eastern Europe. This pattern is consistent with warming temperatures in the spring, which accelerate biological activity and trigger earlier growth cycles. However, northern regions show less pronounced changes, and some areas even exhibit slight delays in SOS, likely due to local climatic variability and delayed snowmelt in colder regions.

Similarly, the EOS has shifted later across much of the Europe and the Mediterranean region. These delays indicate an extension of the growing season, particularly in response to milder autumns and prolonged favourable conditions for photosynthesis. This extension is most pronounced in southern regions, while northern Europe displays more mixed patterns, reflecting the interplay of temperature changes and other local factors, such as moisture availability and distance from warm ocean currents. Earlier EOS in western Europe is a result of an earlier summer start and the high intensity of the growing season.

The start of summer has generally advanced in southern, central, and some regions of western Europe, while its termination has been delayed in most of Europe. These combined shifts suggest a broadening of the summer season, characterized by prolonged periods of higher temperatures. This trend aligns with observed increases in heat extremes, particularly in southern Europe, where prolonged dry spells are becoming more common. The exception is part of western and northern Europe in which both the start and end were significantly earlier towards 2020 with an extended duration of spring and autumn (Fig. 3).

In contrast, winter transitions exhibit a pattern of shortening. The onset of winter is delayed across northern, central and eastern Europe, suggesting a shift towards milder and shorter winter seasons. Meanwhile, the end of winter occurs earlier in most of the Europe and Mediterranean region reflecting warmer winters and earlier SOS. These changes are particularly significant in regions traditionally experiencing longer, harsher winters, where such shifts may have profound ecological and hydrological impacts.

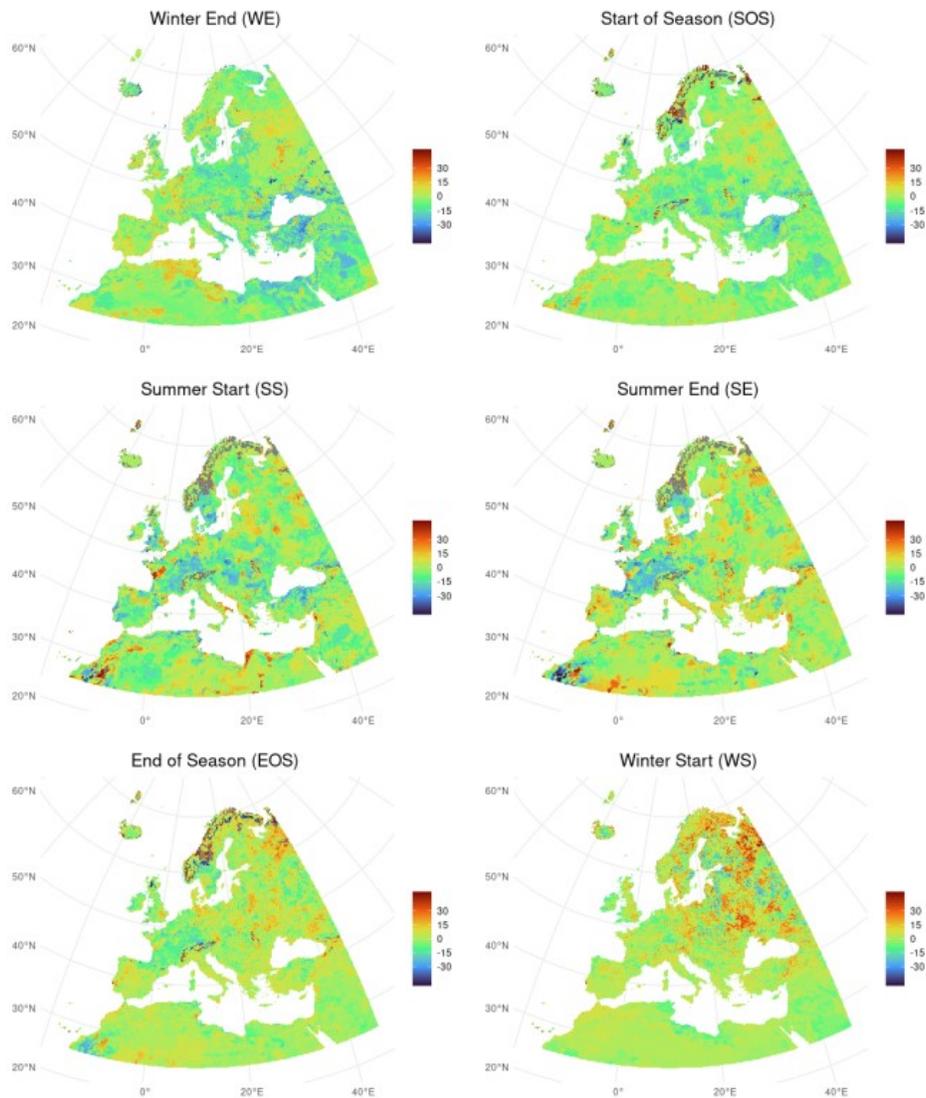

**Figure 7** Shift in onset of seasons over the Euro-Mediterranean region for 1991-2020 ERA5-Land climatology between first (1991-2000) and last (2010-2020) decades.

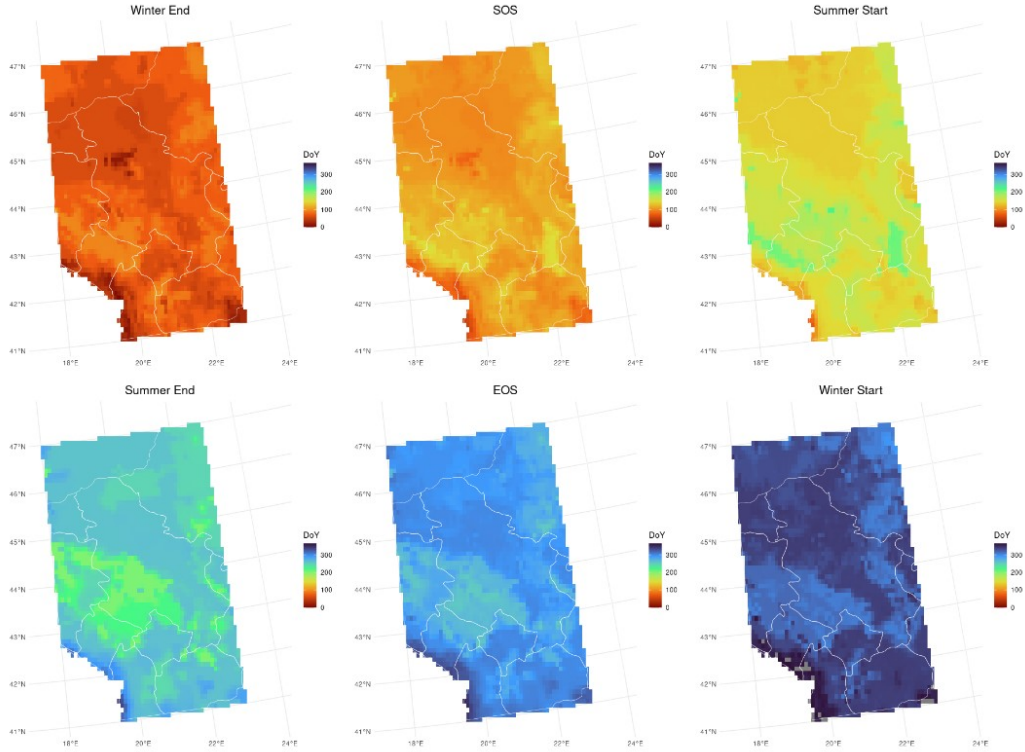

**Figure 8** Onset of seasons over the SE Europe for 1991-2020 ERA5-Land climatology

Finer details of the onset of seasons in SE Europe (Fig. 8) depict later winter ends earlier start on mountain areas and the opposite trend in the rest of the region, particularly on Adriatic Sea coast and the urban heat island of the city of Belgrade. A region can be traced with specifically later winter end and early SOS and summer start as well as late summer end, EOS, and winter start. It starts from the northern part of the region (Central Europe) containing warm air of Pannonian region, for the better part of the year, which can flow through wide river valleys to the south connecting with warm air coming from the Mediterranean sea through Vardar river valley. It forms unique seasonal patterns on small scales picturing a gradient season transition between central Europe and the Mediterranean region.

## 4 Discussion

There is a growing public and expert awareness about shifts in seasonal onset and duration and their impact on all aspects of life and society. It is also evident that no classification of seasons fully supports evidence of seasonal changes from a wide range of natural phenomena (plants and insect phenology, bird migration) and across a wide range of geographic locations. Although many studies address seasons and seasonal transitions, most focus on defining one or two seasons, often neglecting comprehensive methodologies for determining all four seasons. A few studies offer such methodologies, including approaches based on a box-averaged index of continuous raw daily pressure anomalies (Cassou and Cattiaux, 2016) or temperature thresholds set at the 75$^{th}$ percentile of average temperature for summer and 25$^{th}$ percentile for winter (Wang et al., 2021). While the first method is rather complex, requiring significant computational expertise and extensive data sources, the second relies on historical thresholds, defining only summer and winter while treating autumn and spring as transition periods.

In contrast, this study presents a new classification of seasons introducing the normalized daily temperature range (DTRT) as a seasonality index. The results demonstrate that the seasonality index captures seasonal transitions and aligns well with phenology markers such as NDVI, LAI, and fPAR, as well as with ground-based observations of phenological events. The consistency of these findings across diverse geographic regions, including boreal, temperate, and deciduous forests as well as agricultural systems such as crops and orchards (Lalic et al., 2022), underscores the robustness and universality of the proposed index. Unlike traditional definitions, the proposed methodology accommodates atmospheric and vegetative dynamics variations. The seasonality index provides several distinct advantages over existing classification methods. Its simplicity and universality allow it to be applied across diverse climatic regions. The index's independence from vegetation presence makes it particularly valuable in regions with sparse or no vegetation, where traditional phenology-based metrics may not be applicable. Additionally, the DTRT captures temperature variability and energy partitioning associated with seasonal transitions. Its ability to identify fine energy and humidity flux divergence through inflection points and extreme values of the DTRT time series (Lalic et al., 2022) highlights its potential for practical applications in agriculture, forestry, and tourism.

The findings align with earlier studies on changing seasonal dynamics. For instance, earlier SOS trends obtained in this study are consistent with phenology-based trends in Europe (Menzel et al., 2006; Thackeray et al., 2010; Settele et al., 2014). Additionally, across muchof western Europe, our method places  SOS before 100 DOY, aligningwith Cassou and Cattiaux's (2016) definition of summer as the warm part of the year, which begins at 95 DOY.

Our results for spring start (WE) and SOS are in accordance with observed changes in bird's migration and similar seasonal phenomena in some regions that shifted to an earlier date by more than a month (Sparks and Metzel, 2002; Thomson, 2009; Parmesan, 2006; Schwartz et al., 2006). Longer spring and autumn is not necessary equal to longer period between SOS and EOS. Several studies (Cleland et al., 2006; Hollister et al., 2005; Post et al., 2008; Sherry et al., 2007) demonstrate that many species often shorten their life cycle in response to warming. In vegetated areas, the start and end of growing season is clearly identifiable in temperature and humidity time series (see for example Fitzjarrald et al., 2001; Lalic et al., 2022) often appearing as inflection point. A shortened growing season of dominant plant species can affect the timing of this inflection point (EOS timing in our study) even if temperatures are still "summery" high. This shift marks a summer-to-autumn transition and, with a warming climate, can lead to prolonged autumn, as observed in some parts of western Europe (see Figs. 4 and 7).

An earlier summer start in England is in accordance with the findings of Kirbyshire and Bigg (2010), who stated that "summer in England has been advancing since the mid-1950s. Additionally, our results indicating a continual decline of summer and advance of autumn in some regions of northern Europe, particularly the Kola Peninsula (Russia), are in agreement with results presented by Kozlov and Berlina's (2002). The prolonged summer and shorter winter durations obtained in the Euro-Mediterranean region correspond with findings by Wang et al. (2021) and Pena-Ortiz et al. (2015). Our results that indicate the earlier end of summer in some parts of western Europe are in accordance with results presented by Pena-Ortiz et al. (2015) for 1979-2012. More exerted spatial diversity of our results is a consequence of data spatial resolution. Namely, the E-OBS data used in the earlier study has a 0.5° x 0.5° grid (approximately 0.5° = 55.5 km on midlatitudes), while our ERA5-Land data has a 9 km resolution. Finally, let us note that Wang et al. (2021) anticipated that, by 2100, spring and summer will start about 30 days earlier, while autumn and winter will start half month later. According to our results, it already happened in many regions as climate change accelerates.

The main limitation of this study is that there is a lack of a universally accepted benchmark for validation of the proposed classification system; even the comparison with phenology markers and existing studies provides indirect validation. Further research will take several directions: i) challenge the presence and duration of transient periods with outstanding importance for agriculture and tourism; ii) expand analysis to global scales with local analysis for specific climate zones; iii) test its capacity as a method for phenological gap filling; iv) compare seasonality index and insect phenology time series; v) analyze duration of seasons in urban areas using seasonality index; vi) compare seasonality of seasonal flu and variation of seasonality index.

**Conclusion**

Seasons and seasonal changes are complex phenomena. They are result of the unique interplay of atmospheric circulation and different warming rates combined with local factors such as altitude, type of vegetation or vicinity of water bodies and ocean currents.

We developed a classification of seasons that goes in line with human anticipation of seasons. We used vegetation as a nature-based sensor of seasonal transitions. We confirmed capacity of our seasonality index to encompass all factors contributing to seasonal variations and offer clear and simple seasonal metrics that recognizes seasonal differences on small scales even in tropical and subtropical regions.

The spatial variability of seasonal shifts underscores the complexity of climate change impacts within the Euro-Mediterranean region. Southern Europe and the Mediterranean region emerge as hotspots for prolonged summers and delayed winters, while central Europe shows more significant changes in the timing of the growing season. These findings have far-reaching implications for agriculture, water resources management, and ecosystem dynamics. This classification is not only more accurate but also represents a powerful tool in a wide range of disciplines and industries to find solutions to current issues caused by climate change.

**Funding**

This research is supported by the European Union (Grant Agreement No. 101136578). The views and opinions expressed in this publication are solely those of the author(s) and do not necessarily represent the official position of the European Union or the European Commission. Neither the European Union nor the granting authority is responsible for any use that may be made of the information contained herein. Additional support was provided by the Ministry of Science, Technological Development, and Innovation of the Republic of Serbia through two Grant Agreements with the University of Novi Sad, Faculty of Agriculture (No. 451-03-66/2024-03/200117, dated February 5, 2024).

**Data availability**

All datasets used in this research are openly accessible and available online upon registration.